\pdfoutput=1

\documentclass[11pt,twoside,a4paper,cmspaper,final,collab]{cms-tdr}
\def\svnVersion{42598}\def\cmsCernNo{2011-012}\def\cmsCernDate{\today}\def\cmsMessage{Submitted to Physics Letters B}

\begin{document}\cmsNoteHeader{EXO-10-015}

\hyphenation{had-ron-i-za-tion}
\hyphenation{cal-or-i-me-ter}
\hyphenation{de-vices}
\RCS$Revision: 42468 $
\RCS$HeadURL: svn+ssh://alverson@svn.cern.ch/reps/tdr2/papers/EXO-10-015/trunk/EXO-10-015.tex $
\RCS$Id: EXO-10-015.tex 42468 2011-02-28 14:42:44Z hoepfner $

\providecommand\Wprime{\ensuremath{\mathrm{W'}}\xspace}
\providecommand\Zprime{\ensuremath{\mathrm{Z'}}\xspace}
\providecommand\MT{\ensuremath{M_\mathrm{T}} }
\providecommand\invpb{$\mathrm{~pb}^{-1}\,$}
\providecommand\WprimeMuNu{$\Wprime \to \mu \cPgn$}
\providecommand\WMuNu{$\PW \to \mu \cPgn$}

\cmsNoteHeader{EXO-10-015} 
\title{Search for a $\mathrm{W'}\ $ boson decaying to a muon and a neutrino in pp collisions at $\sqrt{s} = 7$ TeV}

\author[cern]{The CMS Collaboration}

\date{\today}

\abstract{
A new heavy gauge boson, \Wprime, decaying to a muon and a neutrino, is searched
for in pp collisions at a centre-of-mass of 7 TeV. The data, collected with the
CMS detector at the LHC, correspond to
an integrated luminosity of 36\invpb.
No
significant excess of events above the standard model expectation is
found in the transverse mass distribution of the muon-neutrino system.
Masses
below 1.40\TeV are excluded at the 95\%
confidence level for a sequential standard-model-like \Wprime.
The \Wprime mass lower limit increases to 1.58\TeV when the present
analysis is combined with the CMS result for
the electron channel.
}

\hypersetup{%
pdfauthor={CMS Collaboration},%
pdftitle={Search for a W' boson decaying to a muon and a neutrino in pp collisions at sqrt(s) = 7 TeV},%
pdfsubject={CMS},%
pdfkeywords={CMS, physics, software, computing}}

\maketitle 

New heavy gauge bosons, generally indicated as \Zprime and \Wprime, are predicted
in various extensions of the standard model (SM).
Such extensions include
Left-Right Symmetric Models~\cite{PhysRevD.11.566,PhysRevD.12.1502,PhysRevD.10.275},
Compositeness models~\cite{compositeness} and  Little Higgs models~\cite{little-higgs}.
The search for a \Wprime is usually performed in the context
of the benchmark model of Ref.~\cite{reference-model}, where
the \Wprime\ boson is considered a heavy analogue of the SM $\PW$ boson with
the same left-handed fermionic couplings.
Thus the \Wprime decay modes and
branching fractions are similar to those of the $\PW$ boson, with the
notable exception of the ${\mathrm t\bar{\mathrm b}}$ channel, which opens up for
\Wprime masses above 180~\GeV.
No interaction with
the SM gauge bosons or with other heavy gauge bosons such as
\Zprime is assumed.
In this context, CDF~\cite{cdf-limit} and D0~\cite{d0-limit}
searched for a \Wprime boson in the decay to an electron and a neutrino,
and excluded \Wprime masses below 1.1\TeV at the 95\% confidence
level (C.L.).
Recently, a search in this decay channel by CMS
extended the lower limit on the \Wprime mass to 1.36\TeV~\cite{cms-electron-limit}.

In this letter, the \Wprime decay to a muon and a neutrino
with an assumed branching fraction of 8.5\% (for all \Wprime masses)
is investigated and combined with a similar search in the electron channel~\cite{cms-electron-limit}.
The data sample,
collected in 2010 by the CMS detector with pp collisions
delivered by the LHC at centre-of-mass energy of 7\TeV, corresponds to 36\invpb.

A more detailed description of CMS can be found elsewhere~\cite{cms}.
The central feature of the CMS apparatus is a superconducting solenoid,
of 6~m internal diameter, providing a magnetic field of 3.8~T. Within the field
volume are the silicon pixel and strip tracker, the crystal electromagnetic calorimeter
and the brass/scintillator hadron calorimeter. Muons are measured
in gas-ionization detectors embedded in the steel return yoke.
Both barrel and endcap regions are instrumented with four muon stations combining
high precision tracking detectors  (drift tubes in the barrel and cathode strip chambers in the endcaps)
with resistive plate chambers for triggering as well as contributing to the tracking. The muon transverse momentum, \pt,
is measured through bending of its track when passing the magnetized return yoke. Each station consists
of a multi-layer chamber, twelve and six layers for the drift and the
cathode strip chambers, respectively. All muon stations contribute to the first level trigger
and identify the bunch crossing from which the muon originated.
A cylindrical coordinate system about the beam axis is used, in which the polar angle $\theta$ is measured
with respect to the counterclockwise beam direction, and $\eta$ is
the
pseudo-rapidity defined by $\eta = -\ln \tan \theta /2$.

Each muon track is matched to a tracker track, measured in the silicon tracker.
A global track fit is performed~\cite{CFT-09-014}, and the resulting
global muon \pt resolution amounts to 1 to 10\% for \pt values up to 1\TeV.
The inner tracker measures charged particle trajectories within the
pseudorapidity range $|\eta| < 2.5$.
It provides an impact parameter resolution of about 15~$\mu$m and a transverse momentum
\pt resolution of about 1.5\,\% for \pt=100\GeV particles.

The primary sources of background to the \WprimeMuNu\ search include standard model \WMuNu\
decays, QCD multijet events, $\mathrm t\bar{\mathrm t}$, Drell-Yan events and cosmic ray muons. 
Diboson
processes (WW, WZ, ZZ) with decays to electrons, muons or taus are also
considered. Monte Carlo (MC) simulation is used to obtain
event samples that are employed to evaluate signal and background efficiencies.
The generated events are processed through the full CMS
\GEANTfour~\cite{Agostinelli:2002hh, Allison:2006ve} based detector
simulation, trigger emulation, and event reconstruction chain. The
\Wprime signal sample is produced with the \PYTHIA 6.409
generator~\cite{Sjostrand:2006za} and the CTEQ6L parton distribution
functions (PDF)~\cite{cteq6}.  Events are generated for \Wprime masses ranging between
0.6\TeV and 1.5~\TeV in steps of 100~\GeV and for a \Wprime mass of
2~\TeV.
Next-to-next-to-leading order (NNLO) corrections
to the \Wprime production cross sections, called  
$k$-factors, are available~\cite{Hamberg:1991p2052,Hamberg:2002p2053}
and amount to
1.32 for $m_{\Wprime}$=0.6\TeV to 1.26 for ${m_\Wprime}$=2\TeV.

The samples for the electroweak background processes \WMuNu\ and
${\rm Z \to} \mu^{+} \mu^{-}$ are produced with
{{\textsc{powheg}}\xspace}~\cite{Alioli:2008gx, Nason:2004rx,
Frixione:2007vw} interfaced with \PYTHIA for showering and
hadronization. Samples of $\PW \to  \mu \nu$, produced with \PYTHIA
with and without the simulation of pile-up effects, are used for
cross-checks. The \PYTHIA generator
is also used for the production of $\PW \to \tau \nu$,
${\rm Z \to} \tau^{+} \tau^{-}$, the diboson (ZZ, WZ, WW) samples, and QCD
multijet events. For $\mathrm t\bar{\mathrm t}$ events, the \MADGRAPH~\cite{Alwall:2007st} generator is used
in combination with \PYTHIA for showering and hadronization. Most
background processes are normalized to the integrated luminosity with NNLO
cross section calculations. However, for $\mathrm t\bar{\mathrm t}$ and QCD multijet,
the NLO and LO cross sections are used, respectively.

Candidate events with at
least one high-\pt muon in the pseudorapidity
range $|\eta|< 2.1$ are selected with a set of
single-muon triggers.
Only global muons reconstructed offline
with \mbox{\pt $>$ 25\GeV} in the range
$|\eta|< 2.1$ are used in the analysis; the global muon track is
required to have  at least eleven hits in the silicon tracker and at least
one hit in the pixel detector.
The global track is also required to satisfy $\chi^2/N_{\rm dof} < 10$
and to have at least two
matching track segments in different muon stations. Since the segments have
multiple hits and are typically found in different muon detectors
separated by thick layers of iron, this requirement significantly
reduces the amount of hadronic punch-through.
The transverse impact parameter $|d_0|$ of a muon track with respect to the
beam spot is required to be
less than 0.02~cm, in order to reduce the cosmic muon background.
Furthermore, the muon is required to be isolated with a combined relative
isolation of less than 0.15.  The latter variable is
defined as the ratio between the addition (within a $\Delta R \equiv \sqrt
{\Delta \phi^2 + \Delta \eta^2} < 0.3$ cone around the muon direction) of the deposited
transverse energy in the calorimeters and the scalar sum
of the transverse momenta of all tracks that
originate from the interaction vertex, and
the muon \pt.
An additional requirement that there be no
more than one muon in the event with \mbox{\pt $>$ 25\GeV} is used to
reduce the Z and Drell-Yan
and cosmic ray muon backgrounds.

\enlargethispage{5mm}

Muon reconstruction, identification, and selection efficiencies, along with
their uncertainties, are determined  from  Z$\to\mu^{+}\mu^{-}$
decays using tag-and-probe techniques.
One lepton candidate, called the ``tag'', satisfies
the trigger criteria and all identification and isolation requirements. The other lepton
candidate, called the ``probe'',
is used to determine the efficiency of specific criteria under study.

The combined muon identification efficiency is measured to be 95\%,
including efficiencies for the muon reconstruction, the muon
selection requirements,
the isolation, and
the inner track measured by the silicon strip tracker.
The value for the combined efficiency is very similar in data and simulation, with the
data/MC ratio being 99\%.
The trigger efficiency was studied with two complementary methods,
the first one using tag-and-probe in dimuon events and the second one using a sample of jet-triggered data, which
result in the trigger efficiencies of 92\% and 91\%, respectively, differing in data and simulation
by 4\%.
Muons from a \Wprime would have higher momenta than those used in the tag-and-probe studies.
So far, only muons up to \pt~=~240\GeV from pp collisions have been recorded and efficiency
studies for O(500)\GeV muons
are done with simulated \Wprime samples.
The combined
efficiency does not depend on \pt despite the fact that the showering probability in
the iron yoke increases with the muon energy. This assumption
has been checked with cosmic muons up to 1\TeV~\cite{CFT-09-014}
and is a consequence
of the redundancy of the muon system.
\begin{figure*}[htb!]
\begin{center}
\includegraphics[width=0.8\textwidth]{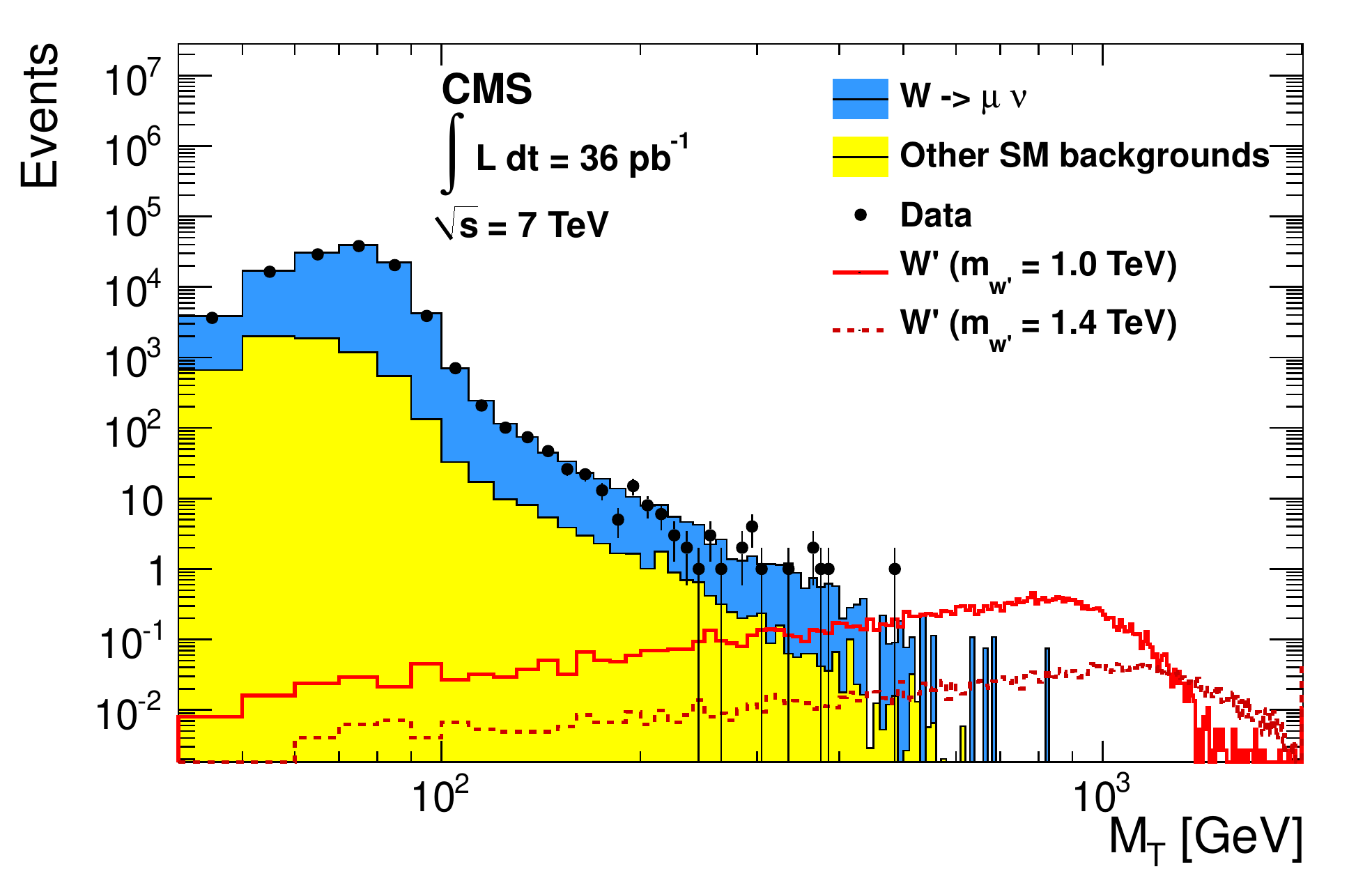}
\end{center}
\caption{The \MT distribution after all the selection steps, in the data and
in the simulation. A \Wprime signal with two different hypothetical masses
is shown.
}
\label{fig:results-MT}
\end{figure*}

The neutrino from a potential \Wprime signal is not detected, but gives rise to missing
transverse energy (\ETmiss) in the detector, which is
calculated using the
particle flow technique~\cite{PFT-09-001}. The technique aims at
reconstructing a complete, unique list of particles in each event using
all the components of the CMS detector: muons, electrons, photons, and
charged and neutral hadrons are all reconstructed
individually.
The \ETmiss for the event is given
by the negative sum of the \pt of all the reconstructed particles in the event.
The \Wprime transverse mass, \MT, is thereafter calculated as:
\begin{equation*}
\MT = \sqrt{2 \cdot \PT  \cdot \MET \cdot (1 - \cos \Delta \phi_{\mu,\nu}) }
\end{equation*}
where $\Delta \phi_{\mu,\nu}$ is the opening azimuthal angle between the muon and the direction of \MET measured in radians.

The two-body decay kinematics is exploited
to further select events with signal-like
topology where the muon and \ETmiss are expected to be nearly
back-to-back in the transverse plane and also balanced in transverse energy.
A selection on the ratio of the muon \pt and \ETmiss is then applied,
$0.4 < \PT/{E_{\mathrm{T}}^{\text{miss}}\xspace} < 1.5$. Further,
the angular difference is required to be
$\Delta \phi_{\mu,\nu} > 2.5$.
After this selection,
the \Wprime signal efficiency for the explored \Wprime mass range
is found to be between 79\% and 82.5\% within
the muon acceptance of $|\eta|< 2.1$.

Estimated SM backgrounds, based on MC simulations,
are shown in Fig.~\ref{fig:results-MT} separately for \PW\ bosons
and for smaller contributions due to QCD,
$\mathrm t\bar{\mathrm t}$, Drell-Yan, and diboson production.
The dominant background up to high transverse masses
is the \WMuNu ~contribution,
which is difficult to suppress as it also decays to a muon and a neutrino.
The data are also shown in
Fig.~\ref{fig:results-MT}, in agreement
with the SM expectation.

The background
in the signal
region is estimated using the lower $180 <$ \MT$< 350$\GeV side band region
of the high \MT part of the spectrum. A relativistic Breit-Wigner function is used as an
ad-hoc empirical shape to fit the \MT distribution in the
side band, both in the simulation and the
data.
The parameters of the fitting function are then used to calculate the
number of expected background events in the  different bins of \MT outside
the side band. The choice of the side band lower and upper limits is made in
order to minimize the contribution from a hypothetical \Wprime signal
and find a region that gives reliable extrapolations of the background in
the signal region, based on simulation studies.
According to the simulation, 71$\pm$8 events are expected in the side band region for the
combination of all SM backgrounds, for an
integrated luminosity of 36\pbinv. The signal contamination would
be 1.63$\pm$0.07 or 0.17$\pm$0.01 \Wprime events for a mass of 1.0 and 1.4\TeV, respectively.
In the data, this region contains 52 events, consistent with the prediction within the
systematic uncertainties in luminosity, background composition and
theoretical uncertainties.
The difference between the background modeling in the data and
the predicted SM background from simulation in various search windows is considered as systematic
uncertainty in the determination of the expected background events in the
signal region. The robustness of this method has been tested by varying
the binning of the \MT distribution and the interval range defining the sideband region,
and it was confirmed that it does not introduce
any significant systematic uncertainties.

Unlike in the electron channel~\cite{cms-electron-limit}, high-\pt cosmic ray muons
constitute an additional source of background for this analysis.
Their spectrum shows a different \MT dependence from that of the dominant
$\PW$ boson contribution. Cosmic ray muons are identified with a
transverse impact parameter with respect to the beam spot larger than 0.02~cm.
The number of cosmic ray muons expected after this requirement is
determined to be between 0.15$\pm$0.04
for \MT$>$350~GeV and 0.08$\pm$0.03 for \MT$>$600~GeV.

\begin{table*}[htb]
\caption{Minimum \MT requirement for the search windows, along with
the expected numbers of signal ($\mathrm{N_{sig}}$) and background ($\mathrm{N_{bkg}}$) events
and data. Also shown are the
theoretical, expected and observed excluded
cross section $\times$ BR
limit for each \Wprime signal sample, using a Bayesian method as discussed in the text.
}
\label{tab:searchwindow-muon}
\centering
\begin{tabular}{|c|c|c|c|c|c|c|c|} \hline
$\mathrm m_\Wprime$  & $\mathrm{M_T}$ & $\mathrm{N_{sig}}$ &
$\mathrm{N_{bkg}}$ & $\mathrm{N_{data}}$ & ${\sigma \cdot {\rm BR}}$ & Expected & Observed \\

(\GeV)  & (\GeV) & (Events)& (Events) & (Events) & (pb)& limit (pb) & limit (pb) \\ \hline
 600 &  390 &152 $\pm$ 16  & 2.54 $\pm$ 0.68 & 1 & 8.290 & 0.308 & 0.212 \\
 700 &  450 & 78 $\pm$ 8   & 1.54 $\pm$ 0.41 & 1 & 4.264 & 0.267 & 0.227 \\
 800 &  470 & 49 $\pm$ 5   & 1.33 $\pm$ 0.35 & 1 & 2.426 & 0.236 & 0.212 \\
 900 &  500 & 29 $\pm$ 3   & 1.09 $\pm$ 0.29 & 0 & 1.389 & 0.216 & 0.150 \\
1000 &  530 & 18 $\pm$ 1.9 & 0.91 $\pm$ 0.23 & 0 & 0.849 & 0.204 & 0.147 \\
1100 &  590 & 11 $\pm$ 1.2 & 0.65 $\pm$ 0.16 & 0 & 0.516 & 0.193 & 0.151 \\
1200 &  610 & 7.1 $\pm$ 0.7 & 0.59 $\pm$ 0.15 & 0 & 0.334 & 0.188 & 0.150 \\
1300 &  630 & 4.7 $\pm$ 0.5 & 0.54 $\pm$ 0.13 & 0 & 0.214 & 0.180 & 0.146 \\
1400 &  630 & 3.0 $\pm$ 0.3 & 0.54 $\pm$ 0.13 & 0 & 0.141 & 0.175 & 0.139 \\
1500 &  680 & 2.1 $\pm$ 0.2 & 0.42 $\pm$ 0.10 & 0 & 0.094 & 0.175 & 0.146 \\
2000 &  690 & 0.29$\pm$ 0.03 & 0.40 $\pm$ 0.10 & 0 & 0.014 & 0.185 & 0.153 \\
\hline
\end{tabular}
\end{table*}

The numbers of background events expected and observed
are reported in
Table~\ref{tab:searchwindow-muon}.
The uncertainties on the number of background events in
Table~\ref{tab:searchwindow-muon} correspond to the statistical uncertainty
of the side band fit itself, as the background is completely determined
from data.

The number of signal events expected
is evaluated from simulation and
given in the third column of
Table~\ref{tab:searchwindow-muon} for different \Wprime masses
along with the total
uncertainty.

The following sources of systematic uncertainties have been considered.

\begin{itemize}
\item {\bf Muon \pt resolution and muon momentum scale:}
Systematic uncertainties due to the muon \pt resolution and momentum
scale are evaluated from detailed studies of the ${\rm Z}\to\mu^{+}\mu^{-}$
mass distribution~\cite{EWK-10-002} and high-\pt cosmic ray muons~\cite{CFT-09-014}.
In order to estimate the effect on the
number of events expected, the data muon \pt spectrum is scaled
and smeared using the values obtained from those studies, the missing transverse energy is recomputed, and finally a
distorted \MT distribution is obtained.
The fit in the side band is performed again with the new \MT distribution.
From comparison with the background estimation obtained from the
original undistorted sample, an uncertainty in the final number of expected
background events for \MT$>$~500\GeV of approximately 3\% is derived.
For the signal yield, where higher values of \pt should be considered,
a conservative value of about 10\% is assigned as systematic uncertainty due to this effect.

\item {\bf \MET resolution:}
An uncertainty of 10\%~\cite{JME-10-004} is assumed on the hadronic component of the \MET
resolution, and used to smear the $x$ and $y$ components of the
reconstructed \MET. The impact on
the number of \Wprime signal events (averaged over all
masses) in the \MT search window with respect to the unsmeared distribution
is found to be below 1\%.

\item {\bf Muon trigger and identification efficiency:}
An uncertainty close to 4\% on the combined muon identification and trigger
efficiency is considered for the signal yield.

\item {\bf Uncertainty on luminosity:} The uncertainty on the absolute
  value of the integrated luminosity is taken as 11\%~\cite{lumi}.

\end{itemize}

\begin{figure*}[htp!]
\centering
\includegraphics[width=0.95\linewidth]{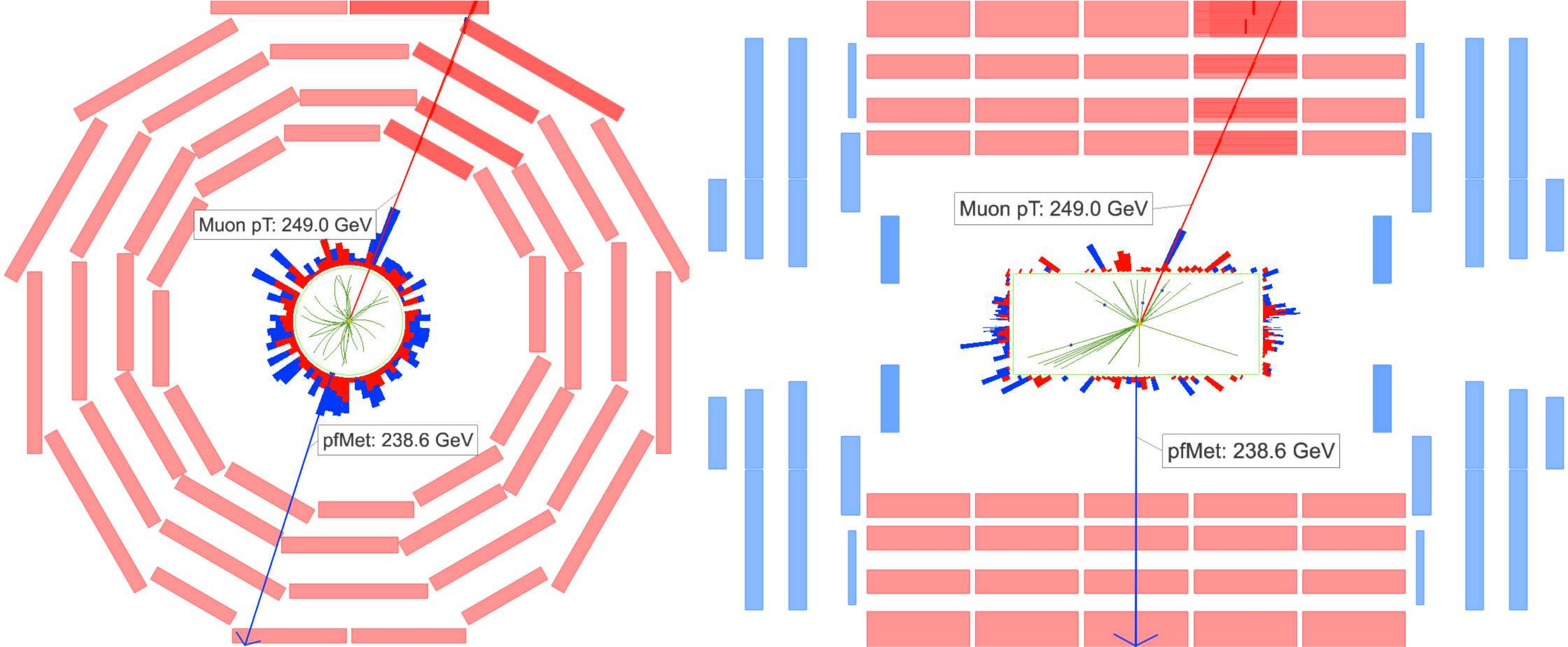}
\caption{Display of the highest-\MT event in transverse view (left)
and longitudinal view (right). The four barrel muon
stations are shown in red, the forward muon stations in blue.
Charged particle tracks as well as the deposited energy per
calorimeter cell are displayed. The muon with \pt=249\GeV
is symbolized by a red line moving upward,
and the \MET=238.6\GeV (``pfMet'')
by a blue line moving downward.
}
\label{fig:evtdisplay}
\end{figure*}

The high \MT region is then used to search for \WprimeMuNu\
which would manifest itself as an excess of events in the TeV region of the \MT
distribution.
No significant excess is observed (Fig.~\ref{fig:results-MT}).
The highest
transverse mass event observed has
\MT = 487~GeV and is displayed in Fig.~\ref{fig:evtdisplay}.

An upper limit is set on the
production cross section times the branching ratio into $\mu\nu$,
$\sigma \cdot
{\rm BR}(W' \to \mu \cPgn)$.
The
limit is derived using the RooStats~\cite{RooStats} interface to BAT~\cite{BAT} to
calculate the limits with the help of Markov Chain Monte Carlo
methods. A flat prior probability distribution is assumed for the
signal cross section. The number of data events above an optimized \MT
threshold are counted and compared to the expected number of
background events.
Systematic uncertainties on the signal efficiency and
background events are treated as nuisance parameters with a log-normal
prior distribution and integrated over. The \MT threshold is optimized
for the best expected limit and defines the lower bound of the search
window for each value of the \Wprime mass. The expected and
observed 95\% C.L. limits for
$\sigma \cdot$ BR are shown in
Table~\ref{tab:searchwindow-muon}.
The value of the theoretical cross
section, shown in
Table~\ref{tab:searchwindow-muon},
is used to translate the excluded cross section into a
\Wprime mass limit. The existence of a
\Wprime with SM-like couplings and masses below 1.40\TeV is excluded at
95\% C.L.
with an expected limit of 1.35\TeV.
Inclusion of \Wprime$\to \tau \cPgn$ decays does not appreciably add to the
acceptance of the \WprimeMuNu\ signal process.

Finally, the results of this search are combined with those of the
${\mathrm W' \to {\rm e} \nu}$ analysis~\cite{cms-electron-limit}.
The muon channel exhibits slightly higher
sensitivity (due to the larger efficiency of about 79-82\%, compared to
64-67\% in the electron channel).
In either channels,
no events are seen at high transverse masses.
Identical NNLO signal
cross sections with the same $k$-factors and
the same PDF uncertainties
are used for both channels under the assumption of lepton universality.
The search windows are optimized individually for each channel based
on the best expected limit. The search windows for both channels
can be found in Table~\ref{tab:comb-limit}.
For each channel the likelihood function is determined and the two likelihood functions are combined.

\begin{figure}[htpb!]
\centering
\includegraphics[width=0.95\textwidth]{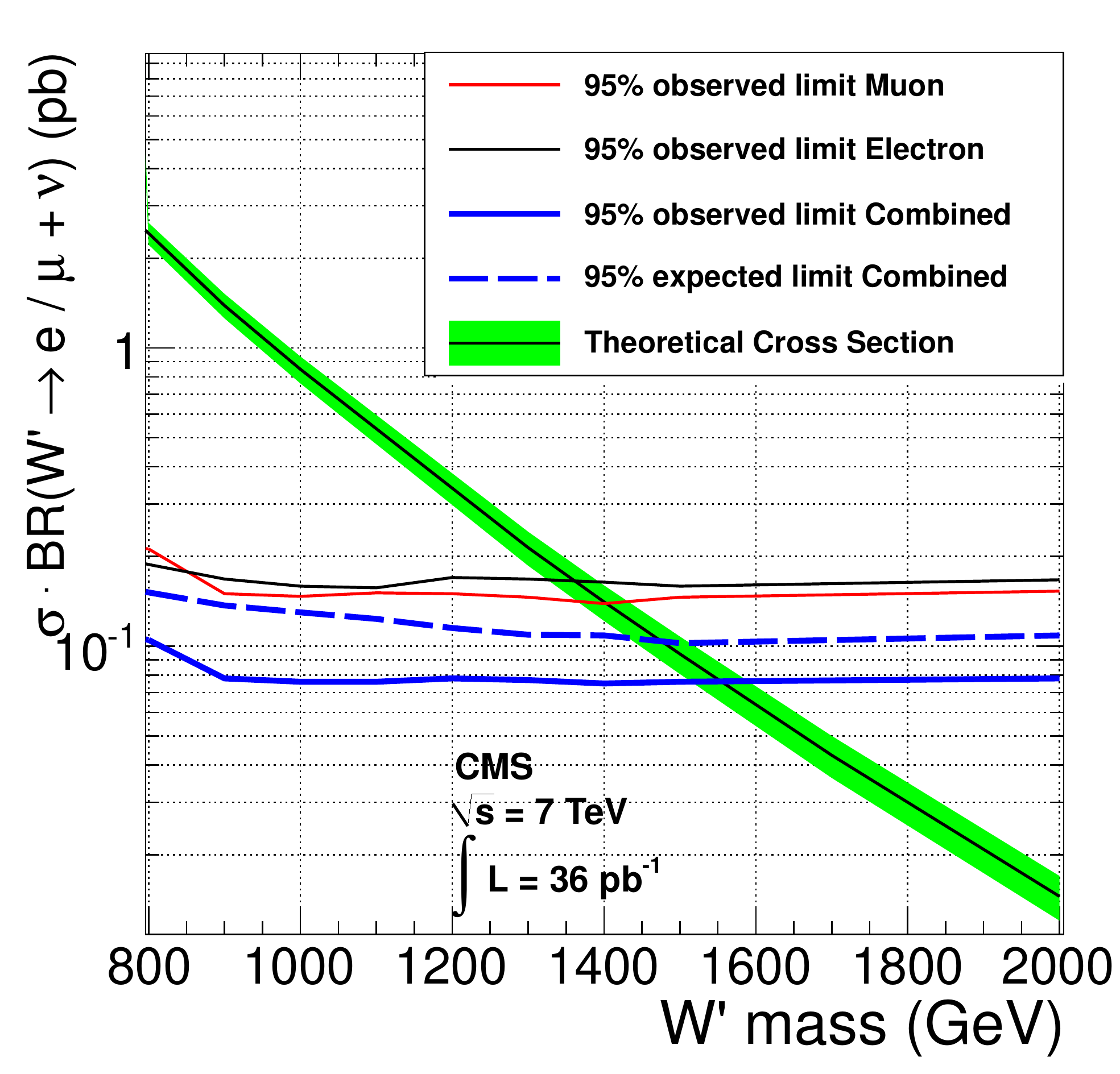}
\caption{Individual limits as observed for the electron (black line) and the muon channel (red line).
Their combination is shown as a solid blue line for the observed limit (expected as dashed blue line),
using a Bayesian technique with only the luminosity uncertainty
correlated.
}
\label{fig:combined-limit}
\end{figure}
The limits for the two individual channels as well as the limit obtained by combining them
are shown in Fig.~\ref{fig:combined-limit}.
The systematic uncertainties for resolution, trigger and lepton identification
efficiencies are assumed to be
fully uncorrelated between both channels.
The uncertainty on the luminosity is taken as fully correlated,
as well as the $k$-factors and PDF uncertainties on the theoretical cross section.
When all background uncertainties are assumed to be fully correlated between the two channels,
the combined limit remains unchanged.
From this combination, a \Wprime with SM-like couplings and with mass
below 1.58~\TeV is excluded at the 95\% confidence level.

\begin{table*}[htb]
\caption{The first two columns show the channel-independent
theoretical NNLO cross section for various \Wprime mass points.
Columns three and five show the individually optimized
minimum \MT thresholds for the electron and muon channels, respectively, along
with the excluded cross section $\times$ BR per channel in columns four and six.
Finally, the right-most column is the combined limit by using a Bayesian method.
\\
}
\label{tab:comb-limit}
\centering
\begin{tabular}{|c|c|c|c|c|c|c|}\hline
$\mathrm m_\Wprime$  & $\sigma \cdot {\rm BR}$ (pb) & $\mathrm{M_T}$ (${\rm
e}$) & Obs. limit & $\mathrm{M_T}$ ($\mu$) &  Obs. limit  &  Obs. limit \\
\GeV   & per channel          & (\GeV)       & electron (pb)      & (\GeV)  & muon (pb) & combined (pb)  \\ \hline
 600 &  8.290 &  400 & 0.289 &  390 & 0.212 & 0.134 \\
 700 &  4.264 &  500 & 0.215 &  450 & 0.227 & 0.116 \\
 800 &  2.426 &  500 & 0.188 &  470 & 0.212 & 0.105 \\
 900 &  1.389 &  500 & 0.168 &  500 & 0.150 & 0.078 \\
1000 &  0.849 &  500 & 0.159 &  530 & 0.147 & 0.076 \\
1100 &  0.516 &  500 & 0.157 &  590 & 0.151 & 0.076 \\
1200 &  0.334 &  650 & 0.170 &  610 & 0.150 & 0.078 \\
1300 &  0.214 &  675 & 0.168 &  630 & 0.146 & 0.077 \\
1400 &  0.141 &  675 & 0.164 &  630 & 0.139 & 0.075 \\
1500 &  0.094 &  675 & 0.159 &  680 & 0.146 & 0.076 \\
2000 &  0.014 &  675 & 0.167 &  690 & 0.153 & 0.078 \\
\hline
\end{tabular}
\end{table*}

In summary, a search for a new heavy gauge boson \Wprime that
decays to a muon and a neutrino
has been performed with 36\pbinv of
data collected by the CMS experiment. No evidence has been found for
\Wprime boson production assuming SM-like couplings
and 95\%
C.L. upper limits have been set on
$\sigma \cdot {\rm BR}(\Wprime \to \mu \nu)$.
Additionally,  a 95\% C.L. lower bound on the  mass of a
\Wprime boson is set at
1.40\TeV.
This lower bound is increased to 1.58\TeV when this analysis is combined
with a similar search for \Wprime $\to {\rm e} \nu$.
This result represents a significant improvement over
previously published limits.

We wish to congratulate our colleagues in the CERN accelerator
departments for the excellent performance of the LHC machine. We thank
the technical and administrative staff at CERN and other CMS
institutes, and acknowledge support from: FMSR (Austria); FNRS and FWO
(Belgium); CNPq, CAPES, FAPERJ, and FAPESP (Brazil); MES (Bulgaria);
CERN; CAS, MoST, and NSFC (China); COLCIENCIAS (Colombia); MSES
(Croatia); RPF (Cyprus); Academy of Sciences and NICPB (Estonia);
Academy of Finland, ME, and HIP (Finland); CEA and CNRS/IN2P3
(France); BMBF, DFG, and HGF (Germany); GSRT (Greece); OTKA and NKTH
(Hungary); DAE and DST (India); IPM (Iran); SFI (Ireland); INFN
(Italy); NRF and WCU (Korea); LAS (Lithuania); CINVESTAV, CONACYT,
SEP, and UASLP-FAI (Mexico); PAEC (Pakistan); SCSR (Poland); FCT
(Portugal); JINR (Armenia, Belarus, Georgia, Ukraine, Uzbekistan); MST
and MAE (Russia); MSTD (Serbia); MICINN and CPAN (Spain); Swiss
Funding Agencies (Switzerland); NSC (Taipei); TUBITAK and TAEK
(Turkey); STFC (United Kingdom); DOE and NSF (USA).

\bibliography{auto_generated}   

\providecommand{\href}[2]{#2}\begingroup\raggedright\begin{thebibliography}{10}%
\makeatletter
\providecommand{\hrefCMSnoop }[0]{\@secondoftwo}%
\makeatother

\bibitem{PhysRevD.11.566}
\hrefCMSnoop {} {R.~N. Mohapatra and J.~C. Pati, ``Left-right gauge symmetry
  and an 'isoconjugate' model of {$CP$} violation'',} \textit{ Phys. Rev.}
  \textbf{ D11} (1975), no.~3, 566.
  \href{http://dx.doi.org/10.1103/PhysRevD.11.566}{\texttt{
  doi:10.1103/PhysRevD.11.566}}.

\bibitem{PhysRevD.12.1502}
\hrefCMSnoop {} {G.~Senjanovic and R.~N. Mohapatra, ``Exact left-right symmetry
  and spontaneous violation of parity'',} \textit{ Phys. Rev.} \textbf{ D12}
  (1975), no.~5, 1502.
  \href{http://dx.doi.org/10.1103/PhysRevD.12.1502}{\texttt{
  doi:10.1103/PhysRevD.12.1502}}.

\bibitem{PhysRevD.10.275}
\hrefCMSnoop {} {J.~C. Pati and A.~Salam, ``Lepton number as the fourth
  `color''',} \textit{ Phys. Rev.} \textbf{ D10} (1974), no.~1, 275.
  \href{http://dx.doi.org/10.1103/PhysRevD.10.275}{\texttt{
  doi:10.1103/PhysRevD.10.275}}.

\bibitem{compositeness}
\hrefCMSnoop {} {E.~Eichten, K.~Lane, and M.~Peskin, ``New tests for quark and
  lepton substructure'',} \textit{ Phys. Rev. Lett} \textbf{ 50} (1983) 811.
  \href{http://dx.doi.org/10.1103/PhysRevLett.50.811}{\texttt{
  doi:10.1103/PhysRevLett.50.811}}.

\bibitem{little-higgs}
\hrefCMSnoop {} {T.~Han {et~al.}, ``Phenomenology of the little {Higgs}
  model'',} \textit{ Phys. Rev.} \textbf{ D67} (2003) 095004,
  \href{http://www.arXiv.org/abs/hep-ph/0301040}{\texttt{
  arXiv:hep-ph/0301040}}.
  \href{http://dx.doi.org/10.1103/PhysRevD.67.095004}{\texttt{
  doi:10.1103/PhysRevD.67.095004}}.

\bibitem{reference-model}
\hrefCMSnoop {} {G.~Altarelli, B.~Mele, and M.~Ruiz-Altaba, ``Searching for New
  Heavy Vector Bosons in ${\rm p\bar{p}}$ Colliders'',} \textit{ Z. Phys.}
  \textbf{ C45} (1989) 109.
  \href{http://dx.doi.org/10.1007/BF01556677}{\texttt{
  doi:10.1007/BF01556677}}.

\bibitem{cdf-limit}
\hrefCMSnoop {} {{CDF Collaboration}, ``Search for a new heavy gauge boson
  {\Wprime} with event signature electron + missing transverse enery in
  $p\bar{p}$ collisions at $\sqrt{s}$ = 1.96 {TeV}'',} \textit{ Phys. Rev.}
  \textbf{ D83} (2011) 031102.
  \href{http://dx.doi.org/10.1103/PhysRevD.83.0311020}{\texttt{
  doi:10.1103/PhysRevD.83.0311020}}.

\bibitem{d0-limit}
\hrefCMSnoop {} {{ D0 Collaboration} Collaboration, ``Search for {\Wprime}
  bosons decaying to an electron and a neutrino with the {\DZERO} detector'',}
  \textit{ Phys. Rev. Lett.} \textbf{ 100} (2008) 031804,
  \href{http://www.arXiv.org/abs/0710.2966}{\texttt{ arXiv:0710.2966}}.
  \href{http://dx.doi.org/10.1103/PhysRevLett.100.031804}{\texttt{
  doi:10.1103/PhysRevLett.100.031804}}.

\bibitem{cms-electron-limit}
\hrefCMSnoop {} {{CMS Collaboration}, ``Search for a heavy gauge boson
  {\Wprime} in the final state with an electron and large missing transverse
  energy in pp collisions at $\sqrt{s}$ = 7{\TeV}''.} article in press in
  \textit{Phys.~Lett.~B}., 2010.
  \href{http://www.arXiv.org/abs/1012.5945v2}{\texttt{ arXiv:1012.5945v2}}.

\bibitem{cms}
\hrefCMSnoop {} {{CMS Collaboration}, ``The {CMS} experiment at the {CERN
  LHC}'',} \textit{ JINST} \textbf{ 3} (2008) S08004.
  \href{http://dx.doi.org/10.1088/1748-0221/3/08/S08004}{\texttt{
  doi:10.1088/1748-0221/3/08/S08004}}.

\bibitem{CFT-09-014}
\hrefCMSnoop {} {{ CMS} Collaboration, ``Performance of {CMS} Muon
  Reconstruction in Cosmic-Ray Events'',} \textit{ JINST} \textbf{ 5} (2010)
  T03022, \href{http://www.arXiv.org/abs/0911.4994}{\texttt{ arXiv:0911.4994}}.
  \href{http://dx.doi.org/10.1088/1748-0221/5/03/T03022}{\texttt{
  doi:10.1088/1748-0221/5/03/T03022}}.

\bibitem{Agostinelli:2002hh}
\hrefCMSnoop {} {{ GEANT4} Collaboration, ``{GEANT4}: A simulation toolkit'',}
  \textit{ Nucl. Instrum. Methods} \textbf{ A506} (2003) 250.
\href{http://dx.doi.org/10.1016/S0168-9002(03)01368-8}{\texttt{
  doi:10.1016/S0168-9002(03)01368-8}}.

\bibitem{Allison:2006ve}
\hrefCMSnoop {} {J.~Allison {et~al.}, ``{Geant4} developments and
  applications'',} \textit{ IEEE Trans. Nucl. Sci.} \textbf{ 53} (2006) 270.
\href{http://dx.doi.org/10.1109/TNS.2006.869826}{\texttt{
  doi:10.1109/TNS.2006.869826}}.

\bibitem{Sjostrand:2006za}
\hrefCMSnoop {} {T.~Sj{\"{o}}strand, S.~Mrenna, and P.~Z. Skands, ``{PYTHIA}
  6.4 Physics and Manual'',} \textit{ JHEP} \textbf{ 05} (2006) 026,
\href{http://www.arXiv.org/abs/hep-ph/0603175}{\texttt{ arXiv:hep-ph/0603175}}.

\bibitem{cteq6}
\hrefCMSnoop {} {J.~Pumplin {et~al.}, ``New Generation of Parton Distributions
  with Uncertainties from Global {QCD} Analysis'',} \textit{ JHEP} \textbf{ 07}
  (2002) 012. \href{http://dx.doi.org/10.1088/1126-6708/2002/07/012}{\texttt{
  doi:10.1088/1126-6708/2002/07/012}}.

\bibitem{Hamberg:1991p2052}
\hrefCMSnoop {} {R.~Hamberg, W.~van Neerven, and T.~Matsuura, ``A complete
  calculation of the order $\alpha_s^2$ correction to the {Drell-Yan}
  {$K$}-factor'',} \textit{ Nucl. Phys.} \textbf{ B359} (1991) 343.
  \href{http://dx.doi.org/10.1016/0550-3213(91)90064-5}{\texttt{
  doi:10.1016/0550-3213(91)90064-5}}.

\bibitem{Hamberg:2002p2053}
\hrefCMSnoop {} {R.~Hamberg, W.~van Neerven, and T.~Matsuura, ``Erratum to a
  complete calculation of the order $\alpha^2_s$ correction to the {Drell-Yan}
  {$K$}-factor (vol {B}359, pg 343, 1991)'',} \textit{ Nucl. Phys.} \textbf{
  B644} (2002) 403.

\bibitem{Alioli:2008gx}
\hrefCMSnoop {} {S.~Alioli {et~al.}, ``{NLO} vector-boson production matched
  with shower in {POWHEG}'',} \textit{ JHEP} \textbf{ 07} (2008) 060,
  \href{http://www.arXiv.org/abs/0805.4802}{\texttt{ arXiv:0805.4802}}.
\href{http://dx.doi.org/10.1088/1126-6708/2008/07/060}{\texttt{
  doi:10.1088/1126-6708/2008/07/060}}.

\bibitem{Nason:2004rx}
\hrefCMSnoop {} {P.~Nason, ``A new method for combining {NLO QCD} with shower
  {Monte Carlo} algorithms'',} \textit{ JHEP} \textbf{ 11} (2004) 040,
  \href{http://www.arXiv.org/abs/hep-ph/0409146}{\texttt{
  arXiv:hep-ph/0409146}}.
\href{http://dx.doi.org/10.1088/1126-6708/2004/11/040}{\texttt{
  doi:10.1088/1126-6708/2004/11/040}}.

\bibitem{Frixione:2007vw}
\hrefCMSnoop {} {S.~Frixione, P.~Nason, and C.~Oleari, ``Matching {NLO QCD}
  computations with Parton Shower simulations: the {POWHEG} method'',} \textit{
  JHEP} \textbf{ 11} (2007) 070,
  \href{http://www.arXiv.org/abs/0709.2092}{\texttt{ arXiv:0709.2092}}.
\href{http://dx.doi.org/10.1088/1126-6708/2007/11/070}{\texttt{
  doi:10.1088/1126-6708/2007/11/070}}.

\bibitem{Alwall:2007st}
\hrefCMSnoop {} {J.~Alwall {et~al.}, ``{MadGraph/MadEvent v4}: The New Web
  Generation'',} \textit{ JHEP} \textbf{ 09} (2007) 028,
  \href{http://www.arXiv.org/abs/0706.2334}{\texttt{ arXiv:0706.2334}}.
\href{http://dx.doi.org/10.1088/1126-6708/2007/09/028}{\texttt{
  doi:10.1088/1126-6708/2007/09/028}}.

\bibitem{PFT-09-001}
\href {http://cdsweb.cern.ch/record/1194487} {{CMS Collaboration},
  ``Particle-Flow Event Reconstruction in {CMS} and Performance for Jets, Taus,
  and {MET}'',} \textit{ CMS Physics Analysis Summary} \textbf{
  CMS-PAS-PFT-09-001} (2009).

\bibitem{EWK-10-002}
\hrefCMSnoop {} {{ CMS} Collaboration, ``Measurements of Inclusive {W} and {Z}
  Cross Sections in pp Collisions at $\sqrt{s}$ = 7 {TeV}'',} \textit{ JHEP}
  \textbf{ 01} (2011).
  \href{http://dx.doi.org/10.1007/JHEP01(2011)080}{\texttt{
  doi:10.1007/JHEP01(2011)080}}.

\bibitem{JME-10-004}
\href {http://cdsweb.cern.ch/record/1279142} {{ CMS} Collaboration, ``Missing
  Transverse Energy Performance in Minimum-Bias and Jet Events from
  Proton-Proton Collisions at $\sqrt{s}$=7{\TeV}'',} \textit{ CMS Physics
  Analysis Summary} \textbf{ CMS-PAS-JME-10-004} (2010).

\bibitem{lumi}
\href {http://cdsweb.cern.ch/record/1279145} {{ CMS} Collaboration,
  ``Measurement of {CMS} Luminosity'',} \textit{ CMS Physics Analysis Summary}
  \textbf{ CMS-PAS-EWK-10-004} (2010).

\bibitem{RooStats}
\hrefCMSnoop {} {L.~Moneta {et~al.}, ``{T}he {R}oo{S}tats {P}roject'',}
  \textit{ {ACAT}2010 {C}onference {P}roceedings} (2010)
  \href{http://www.arXiv.org/abs/1009.1003v1}{\texttt{ arXiv:1009.1003v1}}.

\bibitem{BAT}
\hrefCMSnoop {} {A.~Caldwell, D.~Kollar, and K.~Kr{\"{o}}ninger, ``{BAT} -
  {T}he {B}ayesian {A}nalysis {T}oolkit'',} \textit{ Computer Physics
  Communications} \textbf{ 180} (2009)
  \href{http://www.arXiv.org/abs/0808.2552v1}{\texttt{ arXiv:0808.2552v1}}.

\end{thebibliography}\endgroup

\cleardoublepage\appendix\section{The CMS Collaboration \label{app:collab}}\begin{sloppypar}\hyphenpenalty=5000\widowpenalty=500\clubpenalty=5000\textbf{Yerevan Physics Institute,  Yerevan,  Armenia}\\*[0pt]
S.~Chatrchyan, V.~Khachatryan, A.M.~Sirunyan, A.~Tumasyan
\vskip\cmsinstskip
\textbf{Institut f\"{u}r Hochenergiephysik der OeAW,  Wien,  Austria}\\*[0pt]
W.~Adam, T.~Bergauer, M.~Dragicevic, J.~Er\"{o}, C.~Fabjan, M.~Friedl, R.~Fr\"{u}hwirth, V.M.~Ghete, J.~Hammer\cmsAuthorMark{1}, S.~H\"{a}nsel, M.~Hoch, N.~H\"{o}rmann, J.~Hrubec, M.~Jeitler, G.~Kasieczka, W.~Kiesenhofer, M.~Krammer, D.~Liko, I.~Mikulec, M.~Pernicka, H.~Rohringer, R.~Sch\"{o}fbeck, J.~Strauss, F.~Teischinger, P.~Wagner, W.~Waltenberger, G.~Walzel, E.~Widl, C.-E.~Wulz
% ...
\vskip\cmsinstskip
% ...
1:~~Also at CERN, European Organization for Nuclear Research, Geneva, Switzerland\\
% ...

\end{sloppypar}
\end{document}